\documentclass[prd,reprint,showpacs,showkeys]{revtex4}
\usepackage{amsfonts}
\usepackage{amssymb}
\usepackage{amsmath}
\usepackage{graphicx}
\usepackage[font={footnotesize,it}]{caption}

\begin{document}

\title{Wavy way to the Kerr metric and the quantum nature of its ring
singularity}
\author{O. Gurtug}
\email{ozay.gurtug@emu.edu.tr}
\author{M. Halilsoy}
\email{mustafa.halilsoy@emu.edu.tr}
\affiliation{Department of Physics, Eastern Mediterranean University, G. Magusa, north
Cyprus, Mersin 10, Turkey. }

\begin{abstract}
From inherent non-linearity two gravitational waves, unless they are
unidirectional, fail to satisfy a superposition law. They collide to develop
a new spacetime carrying the imprints of the incoming waves. Same behaviour
is valid also for any massless lightlike field. As a result of the violent
collision process either a naked singularity or a Cauchy horizon (CH)
develops. It was shown by Chandrasekhar and Xanthopoulos (CX) that a
particular class of colliding gravitational waves (CGW) spacetime is locally
isometric to the Kerr metric for rotating black holes. This relation came to
be known as the CX duality. Such a duality can be exploited as an
alternative derivation for the Kerr metric as we do herein. Not each case
gives rise to a CH but those which do are transient to a black hole state
provided stability requirements are met. These classical considerations can
be borrowed to shed light on black hole formation in high energy collisions.
Their questionable stability and many other sophisticated agenda, we admit
that await for a full - fledged quantum gravity. Yet, to add an element of
novelty, a quantum probe is sent in the plane $\theta =\pi /2$ to the naked
ring singularity of Kerr which develops for the overspinning case $(a>M)$ to
test it from a quantum picture. We show that the spatial operator of the
reduced Klein-Gordon equation has a unique self-adjoint extension. As a
result, the classical Kerr`s ring singularity is healed and becomes quantum
regular. Our poetic message of the paper is summarized as

\textit{Let there be light}

\textit{that collide with might}

\textit{to disperse the night}

\textit{and create holes that are white}
\end{abstract}

\maketitle

\section{Introduction}

No doubt the intellectual capacity and motivation boost added to Einstein's
general relativity (GR) and astrophysics/cosmology by the Kerr metric \cite%
{1} has been enormous. It is the rotational extension of the famous
Schwarzschild solution \cite{2} discovered long ago as early as 1916. As the
Schwarzschild metric represents a static black hole solution of Einstein's
equations, the Kerr metric represents its spinning version. Mass and angular
momentum are the physical parameters that characterize the Kerr black hole.
Its original discovery in a pre-computer era such as early 1960's by R. P.
Kerr with the only available logistics of paper, pencil and of course
certain level of intuition and inspiration was almost a miracle. And since
the year 2015 marks the centennial of GR, it should appropriately be in
order to add some minor contributions as an alternative derivation of the
Kerr solution. There are already interesting review articles about the Kerr
metric in the literature \cite{3,4,5,5a}. Our route to Kerr will be entirely
different from our predecessors, it will be a "wavy way" since we
incorporate the collision of gravitational waves.

Being a highly nonlinear theory, rotational effects in GR can not be
encompassed by perturbation techniques. That is, to small order (weak) of
rotation, the metric written down does not reflect the complete properties
of an exact solution. Recall simply that in a non-linear theory,
superposition of two solutions is not a solution. In particular, fast
rotation that highly distorts the spacetime around leads to acausal
connections, closed time-like curves and probably spacetime wormholes. To
say the least, the subject of singularities alone takes an entire life time
for an exhaustive study. For these and many other reasons, Kerr metric made
a mark in the history of science, in particular of relativity and naturally
deserves further comments.

In this modern era of undergoing collision experiments at the Large Hadron
Collider (LHC) at CERN, it is our aim to draw attention that the Kerr metric
emerges locally as a result of colliding gravitational waves (CGW). These
incoming waves consist of an impulsive and shock components in
superposition. This result has been known for at least three decades by the
researchers on colliding waves in GR, however, to inform the rest of the
physics community as well makes one of the principal aim of the present
article.

Collision of gravitational waves with collinear polarization were discovered
first by Khan and Penrose (KP) \cite{6} and Szekeres \cite{sz,7}. The
difference between colliding waves and stationary axially symmetric
spacetimes is that the former has two spacelike while the latter has one
spacelike and one timelike Killing vectors. This implies automatically that
the spacetime of colliding waves is a time dependent, dynamic spacetime.
Extension of the KP solution to non-collinear polarization was achieved by
Nutku and Halil (NH) \cite{8}. The relative polarization angle created a
cross - term in the metric with a naked spacetime singularity weaker than
that of KP solution. By changing the profile of the incoming waves, new
solutions were generated which are listed in the book by Griffiths \cite{9}.
The most interesting among those that makes at the same time the subject
matter of the present article is the solution found by Chandrasekhar and
Xanthopoulos (CX) \cite{10}, (about their formalism see also the work by
Chandrasekhar and Ferrari (CF)\cite{11}). They employed the adjoint solution
of the Ernst equation \cite{12} rather than the standard solution used in
the derivation of the NH solution for colliding waves. The result was
remarkable in the sense that the emerging metric in the interaction region
was locally isometric/transformable to the Kerr metric. Vanishing of the
relative polarization angle between the incoming waves transforms the
resulting metric automatically to the Schwarzschild metric, as expected.
Being locally isometric to the Kerr metric, the CX metric inside the
interaction region of colliding waves is also type-D in Petrov
classification, so that the geodesics and Hamilton-Jacobi equations admit
solutions by separation of variables. On the other hand, the KP/NH geodesics
are not integrable by separation of variables. These are in fact of Petrov
type-I which lacks the separability conditions \cite{13}. Let us add also
that from the universality of the gravitational interaction collision of any
other gauge fields such as electromagnetism or more generally the
non-abelian Yang-Mills fields yield analogous results to gravitational
fields coupled with physical sources. Thus, assuming that the Kerr metric
was not discovered before, it could be discovered, as a result of two CGW
accompanied with a coordinate transformation.

Is there anything deeper that we hover around in all this endeavor ?. If
black holes are considered as matter with their miniscale forms as particles
should there not be a corresponding wave dual to it ?; the spirit of wave -
particle duality. After all, the wave - particle duality lies at the heart
of physics. Based on this analogy, the adjoint solution of CX has both
particle - like and wave - like aspects whereas the normal colliding wave
spacetime (the NH solution) has purely wave - like aspects. Let us add that
Penrose's remark \cite{14}, that every (material) spacetime admits a plane
wave spacetime as a limit is not also independent from this reality. One
step further, in the dynamical Near - Horizon - Geometry, we identify
instead of the plane wave of Penrose the colliding plane waves \cite{15}.

It is shown in the CX duality / isometry that the relative polarization
angle of the waves taking part in the collision becomes proportional to the
angular momentum of the Kerr metric. In other words, the relative
polarization angle of the colliding waves transforms in the isometry to the
rotational degree of the resulting metric. The "equivalent mass" degree of
freedom in the local isometry is derived from the curvature of spacetime.
The lesson to extract from such a result is that in the ultra high energy
collision experiments undergoing at CERN we may search for micro Kerr black
holes \cite{16}. Adding to this the interesting Banados - Silk - West (BSW) 
\cite{17} effect that the center - of - mass (CM) energy in the particle
collisions in the vicinity of the event horizon of the created micro black
holes grow unbounded makes the high energy collision experiments further
important. A severe handicap must be supplemented, however, that the event
horizon of such a black hole is not stable against the slightest
perturbation and decays instantly \cite{18}. At the quantum level this
result can not be detached from the classical picture which says that the
Cauchy horizon developed in colliding waves can not be stable \cite{19,20,21}%
.

We undertake also to investigate from quantum approach the reality of
timelike naked singularities when the rotation parameter dominates the mass
parameter. Unlike the formation of naked singularities in static spacetimes,
naked singularity in the Kerr metric has an exceptional feature. In static
black hole solutions, the naked singularity is central and located at $r=0$.
Hence, all the trajectories hit the singularity once they cross the event
horizon. On the other hand, the naked singularity in the Kerr black hole is
encountered when $r=0$ and $\theta =\pi /2.$ The surface $r=0$ is a disc and
the timelike naked ring singularity forms the boundary of this disc inside
the toroidal ergo region. However, the trajectories approaching this ring
singularity when $\theta \neq \pi /2,$\ automatically miss the singularity,
to encounter regular points. In fact, the scenario for overspinning case
(overextremal, $a>M$) has been questioned by several authors \cite{wd,fl,br}%
, for the sake of weak cosmic censorship. The analysis in these studies have
revealed that by capturing particles, an initial extremal $(a=M)$ black hole
cannot decay into an overspinning case. As a consequence, the possibility of
having an overspinning black hole is unlikely to occur, and excludes the
possibility to have a naked singularity, at least for a restricted class of
scenario. As it was stated in \cite{br}, more detailed studies that cover
all possible scenarios are needed for the resolution of the overspinning
problem.

In this article, the overspinning case of the Kerr's solution that admits
timelike naked singularity is investigated with quantum particles obeying
the Klein-Gordon equation. The Horowitz-Marolf (HM) \cite{hm}, criterion
developed for static spacetimes is extended to stationary spacetimes for a
specific wave mode in which the temporal and spatial parts of the
Klein-Gordon equation are separated. Thus, instead of point particle probe,
the singularity is probed with waves for exploring its quantum nature. Our
finding shows that the ring singularity of Kerr becomes quantum regular.

In view of all these one may conclude that there exist structural
similarities in the mathematical theories between the two seemingly
unrelated topics of GR, namely, the CGW and black holes. This structural
similarity has become more apparent with the discovery of the Ernst
formalism. This is the new formulation of the stationary axially symmetric
gravitational field problem formulated by Ernst in 1968 \cite{12}. Ernst
formalism which leads to the Ernst equation is a cornerstone in obtaining an
exact analytic solution to the field equations of GR. Its symmetries are
described in detail in the book " The Mathematical Theory of Black Holes" by
Chandrasekhar \cite{22}. The formulation of CGW and black holes admit the
same type of Ernst equation. For the sake of completeness, we wish to review
the formalism of Ernst in section II, for the comprehensive analysis which
play a key role in the mathematical structural similarity in theories of
black holes and of CGW. Section III is devoted for exploring this
similarity. First, the derivation of the stationary axisymmetric black hole
solution is briefly explained in Ernst formalism. In the preceding
subsection, the problem of CGW which becomes isometric to the Kerr black
hole is explained in detail. In section IV, the quantum nature of the Kerr's
ring singularity for the overspinning case is investigated within the
framework of quantum mechanics. The paper ends with a conclusion and
discussion in section V.

\section{The Ernst Formalism}

The Ernst formalism introduces an alternative derivation of the field
equations for a uniformly rotating axially symmetric source. This new
formalism involves the use of complex function $\xi $ which is independent
of the azimuthal coordinate. Once $\xi $ is found, a corresponding axially
symmetric solutions of Einstein's or Einstein - Maxwell equations are
constructed. In the present study, our focus will be on the derivation of
the Kerr metric and hence, we shall consider the vacuum Einstein's field
equations.

Our starting point is to adopt the metric known as the Papapetrou - Weyl
form for rotating axially symmetric fields given by%
\begin{equation}
ds^{2}=f^{-1}\left[ e^{2\gamma }\left( dz^{2}+d\rho ^{2}\right) +\rho
^{2}d\varphi ^{2}\right] -f\left( dt-\omega d\varphi \right) ^{2},
\end{equation}%
in which $f=f(\rho ,z),$ $\omega =\omega (\rho ,z)$ \ and $\gamma =\gamma
(\rho ,z).$ Since the metric functions are independent of time coordinate,
it is called stationary axially symmetric. If $\omega =0$, the metric
becomes static axially symmetric. As a requirement of the formalism, the
following Lagrangian density is introduced which involves only the metric
functions $f$ and $\omega ,$%
\begin{equation}
\mathcal{L=-}\frac{1}{2}\rho f^{-2}\overrightarrow{\nabla }f\cdot 
\overrightarrow{\nabla }f+\frac{1}{2}\rho ^{-1}f^{2}\overrightarrow{\nabla }%
\omega \cdot \overrightarrow{\nabla }\omega .
\end{equation}%
By applying the method of variation with respect to $f$ and $\omega ,$ the
following field equations are obtained%
\begin{equation}
f\nabla ^{2}f=\overrightarrow{\nabla }f\cdot \overrightarrow{\nabla }f-\rho
^{-2}f^{4}\overrightarrow{\nabla }\omega \cdot \overrightarrow{\nabla }%
\omega ,
\end{equation}%
\begin{equation}
\overrightarrow{\nabla }\cdot \left( \rho ^{-2}f^{2}\overrightarrow{\nabla }%
\omega \right) =0.
\end{equation}%
Let us add that the mathematical operators gradient $(\overrightarrow{\nabla 
})$, divergence $(\overrightarrow{\nabla }\cdot )$ and Laplacian $(\nabla
^{2})$ are all defined on the flat spacetime, 
\begin{equation}
ds_{0}^{2}=d\rho ^{2}+dz^{2}+\rho ^{2}d\varphi ^{2}.
\end{equation}%
Note also that if $\omega =0,$ Eq.(3) becomes $f\nabla ^{2}f=\left( \nabla
f\right) ^{2}$ and can be integrated easily to find the solutions. The
obtained solution for this particular case is known as Weyl solution.

It is well - known from the vector calculus that for any vector field $%
\overrightarrow{A},$ we have,%
\begin{equation}
\overrightarrow{\nabla }\cdot \left( \overrightarrow{\nabla }\times 
\overrightarrow{A}\right) =0.
\end{equation}%
If we compare Eq.(6) with that of Eq.(4), we have%
\begin{equation}
\rho ^{-2}f^{2}\overrightarrow{\nabla }\omega =\overrightarrow{\nabla }%
\times \overrightarrow{A.}
\end{equation}%
Since the considered spacetime is axially symmetric, it implies independent
of $\varphi $ dependence and as a result,%
\begin{equation}
\omega ,_{\varphi }=0.
\end{equation}%
The above equation imposes the condition that the $\widehat{e}_{\varphi }$
component of the right hand side of the Eq.(7) is zero. Hence, the Eq.(7)
yields,%
\begin{equation}
\omega ,_{\rho }=\rho f^{-2}\left( \rho A_{\varphi ,z}-A_{z,\varphi }\right)
,
\end{equation}%
\begin{equation}
\omega ,_{z}=\rho f^{-2}\left( A_{\rho ,\varphi }\text{ }-\rho A_{\varphi
,\rho }-A_{\varphi }\right) .
\end{equation}%
At this stage we define a new function $F$ such that%
\begin{equation}
F_{,\rho }=A_{\rho },\text{ \ \ \ \ \ \ \ \ \ \ \ \ \ and \ \ \ \ \ \ \ \ \
\ \ \ \ \ }F_{,z}=A_{z},
\end{equation}%
and by introducing a new function $\Phi $ as%
\begin{equation}
\Phi =F_{,\varphi }-\rho A_{\varphi },
\end{equation}%
equations (9) and (10) can be written \ in terms of the new function $\Phi $
as,%
\begin{equation}
\omega ,_{\rho }=-\rho f^{-2}\Phi ,_{z}
\end{equation}%
\begin{equation}
\omega ,_{z}=\rho f^{-2}\Phi ,_{\rho }.
\end{equation}%
Solution to $\omega $ from this pair should also satisfy the integrability
condition%
\begin{equation}
\omega ,_{\rho z}=\omega ,_{z\rho }.
\end{equation}%
This integrability condition implies that the new function $\Phi $ should
also satisfy the same integrability condition%
\begin{equation}
\Phi ,_{\rho z}=\Phi ,_{z\rho }.
\end{equation}%
As a consequence, Eqs.(3) and (4) can be written in terms of new function $%
\Phi $ as,%
\begin{equation}
f\nabla ^{2}f=\left( \overrightarrow{\nabla }f\right) ^{2}-\left( 
\overrightarrow{\nabla }\Phi \right) ^{2},
\end{equation}%
\begin{equation}
\overrightarrow{\nabla }\cdot \left( f^{-2}\overrightarrow{\nabla }\Phi
\right) =0.
\end{equation}%
Once equations (17) and (18) are solved, the other metric functions $\gamma $
and $\omega $ can be obtained by integration. The equations (17) and (18)
can be combined if a new function $\xi $ is defined as 
\begin{equation}
\xi =f+i\Phi ,
\end{equation}%
which yields,%
\begin{equation}
\left( \text{Re}\xi \right) \nabla ^{2}\xi =\left( \nabla \xi \right) ^{2}.
\end{equation}%
This equation is called the Ernst equation for vacuum. Another alternative
way of writing the Ernst equation is to define another function as,%
\begin{equation}
\xi =\frac{E-1}{E+1},
\end{equation}%
leading to,%
\begin{equation}
\left( EE^{\ast }-1\right) \nabla ^{2}E=2E^{\ast }\left( \nabla \xi \right)
^{2},
\end{equation}%
in which "$\ast "$ denotes the complex conjugation. It should be
supplemented once more that the operators are to be evaluated on a base
manifold (5) with $\varphi $ a Killing coordinate.

Let us note that up to this point we have introduced the Ernst formalism
based on the the Weyl - Papapetrou metric and coordinates $\rho $ and $z.$
However, solutions in these coordinates are not much tractable and this
enforces us to consider alternative coordinates. One such system is the
prolate coordinates $x,y$ related to $\rho $ and $z$ by%
\begin{eqnarray}
\rho &=&\sqrt{x^{2}-1}\sqrt{1-y^{2}}, \\
z &=&xy.  \notag
\end{eqnarray}%
By transforming all operators and equations in to the $\left( x,y\right) $
coordinates the Kerr solution follows as the simplest complex solution to
the Ernst equation. We have%
\begin{equation}
E=px-iqy
\end{equation}%
with constants $p$ and $q$ satisfying $p^{2}+q^{2}=1.$ In terms of the
latter coordinates Kerr's metric reads 
\begin{eqnarray}
ds^{2} &=&\frac{\left( p^{2}x^{2}+q^{2}y^{2}-1\right) }{\left( px+1\right)
^{2}+q^{2}y^{2}}\left[ dt-\frac{2q\left( 1-y^{2}\right) \left( px+1\right) }{%
p^{2}x^{2}+q^{2}y^{2}-1}d\varphi \right] ^{2} \\
&&-\frac{\left( px+1\right) ^{2}+q^{2}y^{2}}{p^{2}}\left[ \frac{dx^{2}}{%
x^{2}-1}+\frac{dy^{2}}{1-y^{2}}\right]  \notag \\
&&-\frac{\left( px+1\right) ^{2}+q^{2}y^{2}}{p^{2}x^{2}+q^{2}y^{2}-1}\left(
x^{2}-1\right) \left( 1-y^{2}\right) d\varphi ^{2}.  \notag
\end{eqnarray}%
This form of the Kerr metric is a convinient form to be related with the
colliding wave metric that will be introduced in the next section. From this
form of the Kerr metric by a simple transformation we can transform it into
the Boyer - Lindquist form which is the most familiar form of the metric: we
employ the transformation%
\begin{equation}
px+1=\frac{r}{M},\text{ \ \ \ \ \ }t=t,\text{ \ \ \ \ \ \ }qy=\frac{a}{M}%
\cos \theta ,\text{ \ \ \ \ }\varphi =\varphi ,\text{ \ \ \ \ }p=\frac{\sqrt{%
M^{2}-a^{2}}}{M},\text{ \ \ \ }q=\frac{a}{M}.
\end{equation}%
and obtain%
\begin{eqnarray}
ds^{2} &=&dt^{2}-\left( r^{2}+a^{2}\cos ^{2}\theta \right) \left[ d\theta
^{2}+\frac{dr^{2}}{r^{2}+a^{2}-2Mr}\right] \\
&&-\left( r^{2}+a^{2}\right) \sin ^{2}\theta d\varphi ^{2}-\frac{2Mr}{%
r^{2}+a^{2}\cos ^{2}\theta }\left( dt-a\sin ^{2}\theta d\varphi \right) ^{2}
\notag
\end{eqnarray}%
It can be readily seen that for $a=0$ this reduces to the standard
Schwarzchild line element.

\section{Structural Similarity of the Mathematical Theory of Black Holes and
CGW.}

Spacetimes admitting two Killing fields provide the necessary background for
both the theory of black holes and the theory of CGW. Since we are
interested with stationary axially symmetric metrics; the metric functions
representing black holes are independent of the time $t$ and of the
azimuthal angle $\varphi .$ On the other hand, the metric functions
representing colliding gravitational waves are independent of two spacelike
coordinates $(x^{1},x^{2})$ ranging from $-\infty $ to $+\infty ,$ metric
functions depend on the time $t$ and the spacelike coordinate $x^{3}$ which
is normal to the $(x^{1},x^{2})-planes.$

\subsection{\qquad The Conjugate Solution of the Stationary Axisymmetric
Black Hole Spacetime}

The metric describing the stationary axisymmetric black holes can also be
written in the form adopted by Chandrasekhar \cite{22},%
\begin{equation}
ds^{2}=\sqrt{\Delta \delta }\left[ \chi dt^{2}-\frac{1}{\chi }\left(
d\varphi -\omega dt\right) ^{2}\right] -e^{\mu _{2}+\mu _{3}}\sqrt{\Delta }%
\left[ \left( \frac{d\eta }{\Delta }\right) ^{2}+\left( \frac{d\mu }{\delta }%
\right) ^{2}\right] ,
\end{equation}%
in which%
\begin{equation}
\Delta =\eta ^{2}-1,\text{ \ \ \ \ \ \ \ \ \ }\delta =1-\mu ^{2},\text{ \ \
\ \ \ \ \ \ \ \ \ }\left( \mu =\cos \theta \right) ,
\end{equation}%
where $\eta $ is the radial coordinate, $\chi ,\omega $ and $\mu _{2}+\mu
_{3}$ are the metric functions. The metric function $\omega $ represents the
rotation parameter. As we have experienced from the Ernst formalism, the
whole idea is to solve for the metric functions $\chi $ and $\omega .$ The
remaining metric function $\mu _{2}+\mu _{3},$ can be obtained by simple
integral.

The conjugate metric of the metric (28) can be obtained by applying the
following transformations;%
\begin{equation}
t\text{ }\rightarrow +i\varphi \text{ \ and \ }\varphi \rightarrow -it.
\end{equation}%
This conjugation modifies the metric (28) in the following form 
\begin{equation}
ds^{2}=\sqrt{\Delta \delta }\left[ \widetilde{\chi }dt^{2}-\frac{1}{%
\widetilde{\chi }}\left( d\varphi -\widetilde{\omega }dt\right) ^{2}\right]
-e^{\mu _{2}+\mu _{3}}\sqrt{\Delta }\left[ \left( \frac{d\eta }{\Delta }%
\right) ^{2}+\left( \frac{d\mu }{\delta }\right) ^{2}\right] ,
\end{equation}%
in which,%
\begin{equation}
\widetilde{\chi }=\frac{\chi }{\chi ^{2}-\omega ^{2}},\text{ \ \ \ \ \ \ \ \
\ \ \ \ \ \ \ \ }\widetilde{\omega }=\frac{\omega }{\chi ^{2}-\omega ^{2}}%
\text{\ }
\end{equation}%
Consequently, if the pair of $\left( \chi ,\omega \right) $ is a solution to
the field equations, then the new pair $\left( \widetilde{\chi },\widetilde{%
\omega }\right) $ is also a solution to the same field equations. As a
requirement of the Ernst formalism, in place of $\chi $ and $\omega $, we
define two new functions $\Psi $ and $\Phi $ such that;%
\begin{equation}
\Psi =\frac{\sqrt{\Delta \delta }}{\chi },
\end{equation}%
and $\Phi $ is regarded as the potential for $\omega $ and defined by 
\begin{equation}
\Phi ,_{\eta }=\frac{\delta }{\chi ^{2}}\omega ,_{\mu },\text{ \ \ \ \ \ \ \
\ \ \ \ and \ \ \ \ \ \ \ \ \ \ \ \ \ }\Phi ,_{\mu }=-\frac{\Delta }{\chi
^{2}}\omega ,_{\eta }.
\end{equation}%
One can also write these equations (33) and (34) in terms $\widetilde{\chi }$
and $\widetilde{\omega }$ as;%
\begin{equation}
\widetilde{\Psi }=\frac{\sqrt{\Delta \delta }}{\widetilde{\chi }},
\end{equation}

\begin{equation}
\widetilde{\Phi },_{\eta }=\frac{\delta }{\widetilde{\chi }^{2}}\widetilde{%
\omega },_{\mu },\text{ \ \ \ \ \ \ \ \ \ \ \ \ \ \ and \ \ \ \ \ \ \ \ \ \
\ \ \ }\widetilde{\Phi },_{\mu }=-\frac{\Delta }{\widetilde{\chi }^{2}}%
\widetilde{\omega },_{\eta }.
\end{equation}%
The functions $\Psi ,\Phi $ and $\widetilde{\Psi },\widetilde{\Phi }$ can be
combined into the pairs of complex functions as%
\begin{equation}
Z^{\dag }=\Psi +i\Phi ,\text{ \ \ \ \ \ \ \ \ \ \ \ and \ \ \ \ \ \ \ \ \ \
\ \ }\widetilde{Z}^{\dag }=\widetilde{\Psi }+i\widetilde{\Phi },
\end{equation}%
which is followed by an introduction of new function $E$ as, 
\begin{equation}
E^{\dag }=\frac{Z^{\dag }-1}{Z^{\dag }+1},\text{ \ \ \ \ \ \ \ \ \ \ \ \ and
\ \ \ \ \ \ \ \ \ \ \ }\widetilde{E}^{\dag }=\frac{\widetilde{Z}^{\dag }-1}{%
\widetilde{Z}^{\dag }+1}.
\end{equation}%
These functions admits the following Ernst equation,%
\begin{equation}
\left( 1-\left\vert E\right\vert ^{2}\right) \left\{ \left[ \Delta E_{,\eta }%
\right] ,_{\eta }-\left[ \delta E_{,\mu }\right] ,_{\mu }\right\} =-2E^{\ast
}\left[ \Delta \left( E_{,\eta }\right) ^{2}-\delta \left( E_{,\mu }\right)
^{2}\right] .
\end{equation}%
in which $E$ can be replaced by $E^{\dag }$ or $\widetilde{E}^{\dag }.$

The same solution for the Ernst equation of the last section namely $E=p\eta
-iq\mu ,$ can be employed here in which the prolate coordinates $x$ and $y$
are changed into $\eta $ and $\mu .$ The relation of these coordinates to
the null coordinates $u$ and $v$ will be described in the next section.

\subsection{Formulation of the CGW Problem in Double Null Coordinates}

\begin{figure}[h]
\includegraphics[width=100mm,scale=0.7]{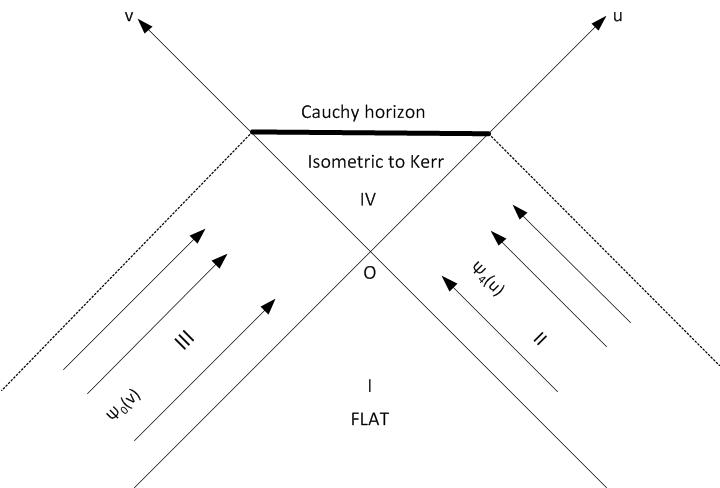}
\caption{The figure depicts collision of two incoming gravitational plane
waves $\Psi _{4}(u)$ (from Region II) and $\Psi _{0}(v)$ (from Region III).
The collision takes place at the origin $O$ ($u=0$ and $v=0$). The post -
collision region (Interaction region or Region IV) is isometric to the Kerr
spacetime. In the diagram, coordinates $x$ and $y$ are orthogonal to the $%
\left( u,v\right) $ plane so that they are suppressed. The hypersurface $%
u^{2}+v^{2}=1,$ locates a Cauchy horizon (CH), not a spacetime singularity,
beyond which the spacetime can be extended analytically to new spacetimes.
It should be added that the stability of the CH against various
perturbations remains as an unsettled dispute. Researchers so far infer
about an unstable CH. }
\label{fig: 1}
\end{figure}

The straightforward technique for CGW is to devise the whole spacetime into
four regions and formulate it as an initial-value problem, as depicted in
figure 1. Usually, Region I, represents the flat Minkowski space in which no
gravitational waves are present. Note that CX interchanges regions I and IV
in their papers. Regions II and III, are the plane symmetric regions which
contains gravitational plane waves that participates in the collision.
Region IV is the interaction region which describes the nonlinear
interaction of gravitational waves in which the resulting spacetime is no
more plane symmetric. The formulation for initial-value problem requires
first to define the waves in the incoming regions, and solve the necessary
field equations in the region of interaction.

However, an alternative method is developed by CF which uses the Ernst
formalism for the formulation of the problem of CGW. In this method, first
the solution in the region of interaction is found and then by making an
extension the waves that participate in the collision is found. This is the
method that we shall employ to drive the Kerr metric in this article.

The adopted line element for the description of the CGW is the Szekeres line
element given by,%
\begin{equation}
ds^{2}=2e^{-M}dudv-e^{-U}\left\{ \left( e^{V}dx^{2}+e^{-V}dy^{2}\right)
\cosh W-2\sinh Wdxdy\right\} ,
\end{equation}%
in which $M=M(u,v),U=U(u,v),V=V(u,v)$ and $W=W(u,v)$ are the metric
functions to be found, all depends on the null coordinates $u$ and $v$ in
the region of interaction. This metric admits two commuting spacelike
Killing vectors $\xi _{1}=\partial _{x}$ and $\xi _{2}=\partial _{y}.$ Note
that in Chandrasekhar's notation $x=x^{1}$ and $y=x^{2}.$ The vacuum
Einstein equations governing the solution to the metric functions are%
\begin{equation}
U_{uv}=U_{u}U_{v},
\end{equation}%
\begin{equation}
2U_{vv}=U_{v}^{2}+W_{v}^{2}+V_{v}^{2}\cosh ^{2}W-2U_{v}M_{v},
\end{equation}%
\begin{equation}
2U_{uu}=U_{u}^{2}+W_{u}^{2}+V_{u}^{2}\cosh ^{2}W-2U_{u}M_{u},
\end{equation}%
\begin{equation}
2V_{uv}=U_{u}V_{v}+U_{v}V_{u}-2\left( V_{u}W_{v}+V_{v}W_{u}\right) \tanh W,
\end{equation}%
\begin{equation}
2M_{uv}=-U_{u}V_{v}+W_{v}W_{u}+V_{u}V_{v}\cosh ^{2}W,
\end{equation}%
\begin{equation}
2W_{uv}=U_{u}W_{v}+U_{v}W_{u}+2V_{u}V_{v}\sinh W\cosh W.
\end{equation}%
Note that these equations, with the exceptions of (42) and (43) follow from
a variational principle of the Lagrangian,%
\begin{equation}
L=e^{-U}\left\{ M_{u}U_{v}+M_{v}U_{u}+U_{u}U_{v}-V_{u}V_{v}\cosh
^{2}W-W_{v}W_{u}\right\} ,
\end{equation}%
where our notation is such that a subscript $u$ $/$ $v$ letter implies
partial derivative. The equations (42) and (43) are simply the integrability
conditions for the other equations. The set of the above field equations can
be solved by employing the Ernst formalism. In doing this, the following
complex valued function is defined,%
\begin{equation}
Z=\chi +iq_{2},
\end{equation}%
where 
\begin{equation}
\chi =\frac{e^{-V}}{\cosh W},\text{ \ \ \ \ \ \ \ \ \ \ \ and \ \ \ \ \ \ \
\ \ \ \ \ \ \ }q_{2}=e^{-V}\tanh W,
\end{equation}%
such that the line element (40) becomes%
\begin{equation}
ds^{2}=2e^{-M}dudv-e^{-U}\left[ \chi dy^{2}+\frac{1}{\chi }\left(
dx-q_{2}dy\right) ^{2}\right] .
\end{equation}%
The field equation (41) can be integrated to give,%
\begin{equation}
e^{-U}=f(u)+g(v),
\end{equation}%
in which $f(u)$ and $g(v)$ are arbitrary functions of their arguments. These
functions can be chosen so that it satisfies the required boundary
conditions on the null boundaries $u=0,$ $v=0$ and hence given by,%
\begin{equation}
e^{-U}=1-u^{2}-v^{2}.
\end{equation}%
Note that within the context of colliding waves the coordinates $u$ and $v$
are to be considered with the Heaviside unit step functions $\theta (u)$ and 
$\theta \left( v\right) $ where 
\begin{equation}
\theta (x)=\left\{ 
\begin{array}{cc}
1, & x>0 \\ 
0, & x<0%
\end{array}%
\right. .
\end{equation}%
That is, we must have the substitutions%
\begin{equation*}
u\rightarrow u\theta \left( u\right) ,\text{ \ \ \ \ \ \ \ \ \ \ \ \ \ and \
\ \ \ \ \ \ \ \ \ \ \ \ \ }v\rightarrow v\theta \left( v\right) .
\end{equation*}%
Obviously the choice, 
\begin{equation*}
e^{-U}=1-u\theta \left( u\right) -v\theta \left( v\right) ,
\end{equation*}%
will give rise to Dirac delta functions in the second derivatives, so this
must be excluded as a possible solution to Eq.(41).

By defining a new set of coordinates \cite{8},%
\begin{equation}
\eta =u\sqrt{1-v^{2}}+v\sqrt{1-u^{2}},\text{ \ \ \ \ \ \ \ \ \ \ \ and \ \ \
\ \ \ \ \ \ \ }\mu =u\sqrt{1-v^{2}}-v\sqrt{1-u^{2}},
\end{equation}%
the metric that describes the collision of gravitational waves in the region
of interaction is transformed to the following form,%
\begin{equation}
ds^{2}=e^{\nu +\mu _{3}}\sqrt{\Delta }\left[ \frac{\left( d\eta \right) ^{2}%
}{\Delta }-\frac{\left( d\mu \right) ^{2}}{\delta }\right] -\sqrt{\Delta
\delta }\left[ \chi \left( dy\right) ^{2}+\frac{1}{\chi }\left(
dx-q_{2}dy\right) ^{2}\right] ,
\end{equation}%
where $\eta $ defines the time from the instant of the collision, $\mu $
defines the distance in the normal direction to the spacelike $(x,y)-planes$
with,%
\begin{equation}
\Delta =1-\eta ^{2},\text{ \ \ \ \ \ \ \ \ \ \ \ \ \ \ and \ \ \ \ \ \ \ \ \
\ \ \ \ \ }\delta =1-\mu ^{2},
\end{equation}%
and $\chi $, $\nu +\mu _{3}$ and $q_{2}$ are the metric functions to be
found. It should be noted that the metric functions $q_{2},$ measures the
variation in the polarization of the gravitational waves. If $q_{2}=0,$ the
gravitational waves are said to be linearly polarized. Note also that in the
transformation $u\rightarrow \sin u$ \ and $v\rightarrow \sin v$, instead of
(54) is the choice employed by CX \cite{11}. In accordance with the latter
choice we have $\eta =\sin \left( u+v\right) $ and $\mu =\sin \left(
u-v\right) $ with possible scalings of the null coordinates, such as $%
u\rightarrow au$ and $v\rightarrow bv$ with $\left( a,b\right) $ constants.

Following the similar steps as was used in the mathematical theory of black
holes \cite{22}, the new parametrization;%
\begin{equation}
\chi +iq_{2}=Z=\frac{1+E}{1-E},
\end{equation}%
one obtains the Ernst equation in terms of $Z$ as%
\begin{equation}
\left( Z+Z^{\ast }\right) \left[ \left( \Delta Z_{,_{\eta }}\right) ,_{\eta
}-\left( \delta Z_{,\mu }\right) ,_{\mu }\right] =2\left[ \Delta \left(
Z_{,_{\eta }}\right) ^{2}-\delta \left( Z_{,_{\mu }}\right) ^{2}\right] ,
\end{equation}%
and in terms of $E$ as,%
\begin{equation}
\left( 1-\left\vert E\right\vert ^{2}\right) \left\{ \left[ \Delta E_{,\eta }%
\right] ,_{\eta }-\left[ \delta E_{,\mu }\right] ,_{\mu }\right\} =-2E^{\ast
}\left[ \Delta \left( E_{,\eta }\right) ^{2}-\delta \left( E_{,\mu }\right)
^{2}\right] ,
\end{equation}%
which is the same equation of (39). The equations related to metric function 
$\nu +\mu _{3},$ can be written in terms of $E$ and given by%
\begin{equation}
\frac{\mu }{\delta }\left( \nu +\mu _{3}\right) ,_{\eta }+\frac{\eta }{%
\Delta }\left( \nu +\mu _{3}\right) ,_{\mu }=-\frac{1}{\chi ^{2}}\left( \chi
,_{\eta }\chi ,_{\mu }+q_{2},_{\eta }q_{2},_{\mu }\right) =-2\frac{E,_{\eta
}E^{\ast },_{\mu }+E^{\ast },_{\eta }E,_{\mu }}{\left( 1-\left\vert
E\right\vert ^{2}\right) ^{2}},
\end{equation}%
and%
\begin{multline}
2\eta \left( \nu +\mu _{3}\right) ,_{\eta }+2\mu \left( \nu +\mu _{3}\right)
,_{\mu }=\frac{3}{\Delta }+\frac{1}{\delta }-\frac{1}{\chi ^{2}}\left\{
\Delta \left[ \left( \chi ,_{\eta }\right) ^{2}+\left( q_{2},_{\eta }\right)
^{2}\right] +\delta \left[ \left( \chi ,_{\mu }\right) ^{2}+\left(
q_{2},_{\mu }\right) ^{2}\right] \right\} \\
=\frac{3}{\Delta }+\frac{1}{\delta }-\frac{4}{\left( 1-\left\vert
E\right\vert ^{2}\right) ^{2}}\left[ \Delta \left\vert E,_{\eta }\right\vert
^{2}+\delta \left\vert E,_{\mu }\right\vert ^{2}\right] .
\end{multline}%
The derivation of the metric function $q_{2}$, is possible from a potential $%
\Phi $, as in the case for the metric function $\omega $ of the black hole
case, we have%
\begin{equation}
\Phi ,_{\eta }=\frac{\delta }{\chi ^{2}}q_{2},_{\mu }\text{ \ \ \ \ \ \ \ \
\ \ and \ \ \ \ \ \ \ \ \ \ \ \ }\Phi ,_{\mu }=\frac{\Delta }{\chi ^{2}}%
q_{2},_{\eta },
\end{equation}%
and defining%
\begin{equation}
Z^{\dag }=\Psi +i\Phi =\frac{1+E^{\dag }}{1-E^{\dag }},
\end{equation}%
in which%
\begin{equation}
\Psi =\frac{\sqrt{\Delta \delta }}{\chi }.
\end{equation}%
With this formalism, the corresponding equation for $Z^{\dag }$ and the
equation for $E^{\dag },$ which is the Ernst equation are obtained
respectively,%
\begin{equation}
\left( Z^{\dag }+\left( Z^{\dag }\right) ^{\ast }\right) \left[ \left(
\Delta \left( Z^{\dag }\right) _{,_{\eta }}^{\ast }\right) ,_{\eta }-\left(
\delta \left( Z^{\dag }\right) _{,\mu }^{\ast }\right) ,_{\mu }\right] =2%
\left[ \Delta \left( \left( Z^{\dag }\right) _{,_{\eta }}^{\ast }\right)
^{2}-\delta \left( \left( Z^{\dag }\right) _{,_{\mu }}^{\ast }\right) ^{2}%
\right] ,
\end{equation}%
\begin{equation}
\left( 1-\left\vert \left( E^{\dag }\right) \right\vert ^{2}\right) \left\{ 
\left[ \Delta \left( E^{\dag }\right) _{,\eta }\right] ,_{\eta }-\left[
\delta \left( E^{\dag }\right) _{,\mu }\right] ,_{\mu }\right\} =-2\left(
E^{\dag }\right) ^{\ast }\left[ \Delta \left( \left( E^{\dag }\right)
_{,\eta }\right) ^{2}-\delta \left( \left( E^{\dag }\right) _{,\mu }\right)
^{2}\right] ,
\end{equation}%
and the equations governing $\nu +\mu _{3},$ can be expressed in terms of $%
E^{\dag }$ as%
\begin{multline}
\frac{1}{\chi ^{2}}\left( \chi ,_{\eta }\chi ,_{\mu }+q_{2},_{\eta
}q_{2},_{\mu }\right) =\left( \ln \chi \right) ,_{\eta }\left( \ln \chi
\right) ,_{\mu }+\frac{\Phi ,_{\eta }\Phi ,_{\mu }}{\Psi ^{2}} \\
=\left( \frac{\eta }{\Delta }+\frac{\Psi ,_{\eta }}{\Psi }\right) \left( 
\frac{\mu }{\delta }+\frac{\Psi ,_{\mu }}{\Psi }\right) +\frac{\Phi ,_{\eta
}\Phi ,_{\mu }}{\Psi ^{2}} \\
=\frac{\mu }{\delta }\left[ \ln \frac{\Psi }{\sqrt[4]{\Delta \delta }}\right]
_{,\eta }+\frac{\eta }{\Delta }\left[ \ln \frac{\Psi }{\sqrt[4]{\Delta
\delta }}\right] _{,\mu }+\frac{\Phi ,_{\eta }\Phi ,_{\mu }+\Psi ,_{\eta
}\Psi ,_{\mu }}{\Psi ^{2}}.
\end{multline}%
Hence, equation (60) transforms to the form%
\begin{multline}
\frac{\mu }{\delta }\left[ \left( \nu +\mu _{3}\right) +\ln \frac{\Psi }{%
\sqrt[4]{\Delta \delta }}\right] _{,\eta }+\frac{\eta }{\Delta }\left[
\left( \nu +\mu _{3}\right) +\ln \frac{\Psi }{\sqrt[4]{\Delta \delta }}%
\right] _{,\mu } \\
=-\frac{\Phi ,_{\eta }\Phi ,_{\mu }+\Psi ,_{\eta }\Psi ,_{\mu }}{\Psi ^{2}}%
=-2\frac{E^{\dag },_{\eta }\left( E^{\dag }\right) ^{\ast },_{\mu }+\left(
E^{\dag }\right) ^{\ast },_{\eta }E^{\dag },_{\mu }}{\left( 1-\left\vert
E^{\dag }\right\vert ^{2}\right) ^{2}},
\end{multline}%
and the equation (61) transforms to the form%
\begin{multline}
2\eta \left[ \left( \nu +\mu _{3}\right) +\ln \frac{\Psi }{\sqrt[4]{\Delta
\delta }}\right] ,_{\eta }+2\mu \left[ \left( \nu +\mu _{3}\right) +\ln 
\frac{\Psi }{\sqrt[4]{\Delta \delta }}\right] ,_{\mu } \\
=\frac{3}{\Delta }+\frac{1}{\delta }-\frac{4}{\left( 1-\left\vert E^{\dag
}\right\vert ^{2}\right) ^{2}}\left[ \Delta \left\vert E^{\dag },_{\eta
}\right\vert ^{2}+\delta \left\vert E^{\dag },_{\mu }\right\vert ^{2}\right]
.
\end{multline}

\subsubsection{Chandrasekhar - Xanthopoulos (CX) Solution}

The fundamental study in the field of colliding gravitational waves is the
KP solution which describes the collision of two impulsive gravitational
waves with parallel polarization. The generalization of KP solution for
nonaligned polarization is given by the NH solution. The latter is rederived
by CF \cite{11} by employing the Ernst formalism and it is shown that, if
the solution to Eq.(59) is taken as 
\begin{equation}
E=p\eta +iq\mu ,\text{ \ \ \ \ \ \ \ \ \ \ \ \ \ with \ \ \ \ \ \ \ \ \ \ \ }%
p^{2}+q^{2}=1,
\end{equation}%
solution to the metric functions leads to the NH solution. Note that
previous choice with $q\rightarrow -q$ for the Ernst equation is admissible
provided we employ the same convention throughout. Note that $q$ measures
the second (or cross) polarization of the waves involved. With $q=0$ $(p=1)$
this solution reduces to the solution of linearly polarized wave problem of
KP. It is important to note that, the NH solution is not of Petrov - type D.
However, if the same function in Eq.(70) is considered for $E^{\dag }$ for
the Ernst equation (66), i.e.%
\begin{equation}
E^{\dag }=p\eta +iq\mu ,\text{ \ \ \ \ \ \ \ \ \ \ \ \ \ with \ \ \ \ \ \ \
\ \ \ \ }p^{2}+q^{2}=1,
\end{equation}%
yields the conjugate expression 
\begin{equation}
Z^{\dag }=\Psi +i\Phi =\frac{1+p\eta +iq\mu }{1-p\eta -iq\mu },
\end{equation}%
from which we can find $\Psi $ and $\Phi $ separately as,%
\begin{equation}
\Psi =\frac{1-p^{2}\eta ^{2}-q^{2}\mu ^{2}}{\left( 1-p\eta \right)
^{2}+q^{2}\mu ^{2}},\text{ \ \ \ \ \ \ \ \ \ \ and \ \ \ \ \ \ \ \ \ \ }\Phi
=\frac{2q\mu }{\left( 1-p\eta \right) ^{2}+q^{2}\mu ^{2}}.
\end{equation}%
The metric function $\chi $ is readily available from Eq.(64) as,%
\begin{equation}
\chi =\sqrt{\Delta \delta }\frac{\left( 1-p\eta \right) ^{2}+q^{2}\mu ^{2}}{%
1-p^{2}\eta ^{2}-q^{2}\mu ^{2}},
\end{equation}%
while $q_{2}$ is obtained from Eq.(62) by simple integration as,%
\begin{equation}
q_{2}=\frac{2q}{p\left( 1+p\right) }-\frac{2q\delta \left( 1-p\eta \right) }{%
p\left( 1-p^{2}\eta ^{2}-q^{2}\mu ^{2}\right) }.
\end{equation}%
The remaining metric function $\nu +\mu _{3}$ can be obtained by using
equations (68) and (69) and the result is%
\begin{equation}
e^{\nu +\mu _{3}}=\frac{\left( 1-p\eta \right) ^{2}+q^{2}\mu ^{2}}{\sqrt{%
\Delta }}.
\end{equation}%
Consequently, the metric that describes the collision of gravitational waves
for the assumed solution of Ernst equation is%
\begin{equation}
ds^{2}=X\left[ \frac{\left( d\eta \right) ^{2}}{\Delta }-\frac{\left( d\mu
\right) ^{2}}{\delta }\right] -\Delta \delta \frac{X}{Y}dy^{2}-\frac{Y}{X}%
\left( dx-q_{2}dy\right) ^{2},
\end{equation}%
where%
\begin{equation}
X=\left( 1-p\eta \right) ^{2}+q^{2}\mu ^{2},\text{ \ \ \ \ \ \ \ \ \ \ \ \ }%
Y=1-\left\vert \left( E^{\dag }\right) \right\vert ^{2}=1-p^{2}\eta
^{2}-q^{2}\mu ^{2}=p^{2}\Delta +q^{2}\delta ,
\end{equation}%
and%
\begin{equation}
q_{2}=\frac{2q}{p\left( 1+p\right) }-\frac{2q\delta \left( 1-p\eta \right) }{%
pY},
\end{equation}%
CX proceeded also with the analytic extension of their solution of colliding
waves. In this formalism different universes are patched together, ending up
with new universes as white holes. Example of more general analytic
extension with more parameters such as charge \cite{kn} and
Newman-Unti-Tamburino (NUT) \cite{nut} also is available in the literature 
\cite{og}

\subsubsection{ Extension of the Interaction Region into Incoming Regions}

In order to find the wave profiles that participate in the collision, the
metric obtained in the interaction region should be extended to the incoming
regions. The approaching wave in one of the plane symmetric Region II $%
\left( u\geq 0,v<0\right) $ is obtained by dropping the $v$ in the metric
(77). This process follows systematically with $v<0,$ since the step
function $\theta \left( v\right) =0$ for $v<0$ and the $v$ dependence
disappears. This is achieved by the substitution $\eta =\mu =\sin \left(
u\theta (u)\right) ,$ so that the metric functions take the form%
\begin{equation}
X(u)=1-2p\sin u+\sin ^{2}u,\text{ \ \ \ \ \ \ \ \ }Y(u)=\Delta =\delta =\cos
^{2}u,
\end{equation}

\begin{equation}
q_{2}(u)=\frac{2q}{\left( 1+p\right) }\left[ \left( 1+p\right) \sin u-1%
\right] .
\end{equation}%
in which as described above $u$ is implied with the step function. Hence,
the metric in Region II in terms of the null coordinate $u$ can be expressed
as%
\begin{equation}
ds^{2}=\frac{2X(u)}{\sqrt{1-u^{2}}}dudv-\left( 1-u^{2}\right) \left[
X(u)dy^{2}+\frac{1}{X(u)}\left( d\widetilde{x}-2q\sin udy\right) ^{2}\right]
,
\end{equation}%
where%
\begin{equation}
\widetilde{x}=x+\frac{2q}{\left( 1+p\right) }y.
\end{equation}%
The plane symmetric metric (82) has a single curvature tensor component
which describes the profile of the incoming gravitational wave given by%
\begin{equation}
\Psi _{4}=-\left( p-iq\right) \delta \left( u\right) -\frac{3\left(
X-2iq\sin u\right) }{X^{4}\sqrt{X^{2}+4q^{2}\sin ^{2}u}}\frac{\left( 1-p\sin
u-iq\sin u\right) ^{3}}{\left( p+iq\right) ^{2}}\theta (u),
\end{equation}%
in which $\delta \left( u\right) $ stands for the Dirac delta function and $%
X=X(u)$ is given (80). Consequently, the incoming wave is a composition of
an impulsive and shock gravitational waves. Similar incoming wave profile $%
\Psi _{0}\left( v\right) $ from the Region III $(u<0,v\geq 0)$ is obtained
by the substitution $\eta =-\mu =\sin \left( v\theta \left( v\right) \right)
,$ which will not be given.

\subsubsection{CX - Duality leading to the Kerr Metric}

The Petrov classification of the metric (77), as is shown by CX is type - D.
Calculations for the Weyl scalar with a proper tetrads reveals the only
nonvanishing scalar as%
\begin{equation}
\Psi _{2}=\frac{1}{2\left( 1-p\eta -iq\mu \right) ^{3}}.
\end{equation}%
The Weyl scalar $\Psi _{2}$, in the terminology of the CGW is interpreted as
the Coulomb component and arises as a result of the non-linear interaction.
The unboundedness of $\Psi _{2}$ indicates the existence of the curvature
singularity. As it was shown in the KP and NH solutions, there is a
curvature singularity in the region of interaction when $u^{2}+v^{2}=1$.
This surface corresponds to $\eta =1,$ for the metric (77). And hence, the
behaviour of $\Psi _{2}$ is finite which indicates Killing - Cauchy horizon
instead of a curvature singularity.

We apply the following transformations to the metric (77) which describes
the interaction region of the collision of impulsive and shock gravitational
waves,

\begin{figure}[h]
\includegraphics[width=100mm,scale=0.7]{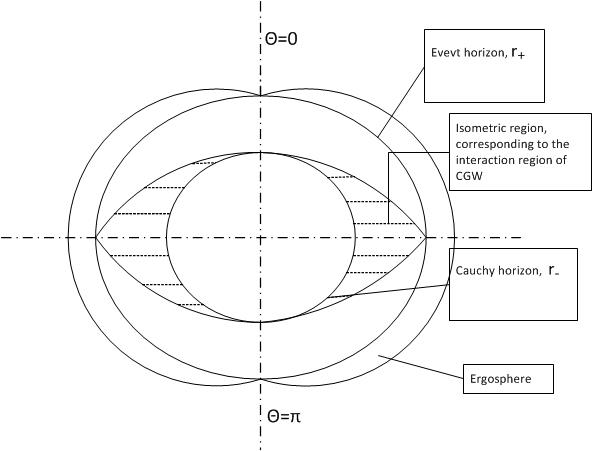}
\caption{The figure represent a partial correspondence between the Kerr
black hole interior and the interaction region of colliding plane waves. The
shaded region correspond to the interaction region of colliding
gravitational plane wave solution (Eq. 72) which is locally isometric to the
part of the region in between the Cauchy $(r_{-})$ and event $\left(
r_{+}\right) $ horizons of the Kerr black hole. The location of the Cauchy
horizon of colliding plane wave $\protect\eta =1,$ corresponds to the inner
horizon $(r_{-})$ of the Kerr black hole located at $r_{-}=M-\protect\sqrt{%
M^{2}-a^{2}}.$ The collision occurs at $\protect\eta =\protect\mu =0,$ which
is equivalent to (from Eq. 81) $r=M $ and $\protect\theta =\protect\pi /2,$
in the corresponding Kerr geometry.}
\label{fig: 2}
\end{figure}

\begin{equation}
t=M\left( x-\frac{2q}{p\left( 1+p\right) }y\right) ,\text{ \ \ \ \ \ \ \ \ \
\ \ }\phi =\frac{M}{\sqrt{M^{2}-a^{2}}}y\text{ \ \ \ \ \ \ \ \ \ \ }\eta
=\pm \frac{\left( M-r\right) }{\sqrt{M^{2}-a^{2}}}\text{ \ \ \ \ \ \ \ \ \ }%
\mu =\cos \theta ,
\end{equation}%
with%
\begin{equation}
p=\pm \frac{\sqrt{M^{2}-a^{2}}}{M},\text{ \ \ \ \ \ \ \ \ \ \ \ \ }q=\pm 
\frac{a}{M},\text{ \ \ \ \ \ and \ \ \ }M^{2}>a^{2}.
\end{equation}%
such that $p^{2}+q^{2}=1.$ We have the correspondence \cite{10}, accordingly

\begin{equation}
1-p\eta =\frac{r}{M},\text{ \ \ \ \ \ \ \ \ \ \ }1-\eta ^{2}=-\frac{\tilde{%
\Delta}}{M^{2}-a^{2}},
\end{equation}%
in which $\tilde{\Delta}$ stands for the horizon function 
\begin{equation}
\tilde{\Delta}=r^{2}-2Mr+a^{2}=\left( r-r_{-}\right) \left( r-r_{+}\right) .
\end{equation}%
We note that this transformation (86-87) is suggestive upon the striking
similarity between (72) and (25) in which the Killing fields $\xi _{x}$ and $%
\xi _{y}$ in one are replaced by the $\xi _{t}$ and $\xi _{\varphi }$ in the
other. An interesting aspect of this transformation is that the infinite
ranged coordinate y is mapped into the compact coordinate $\varphi .$ The
non-orthogonality of Killing vectors in one is provided by $q$ (the relative
polarization of waves) in one while by $a$ (rotation) in the other. It is
natural that in the transformation one ($q$) will be related to the other ($%
a $). These substitutions transform the metric(77) to the following form,%
\begin{equation}
M^{2}ds^{2}=\left( \frac{\tilde{\Delta}-a^{2}\sin ^{2}\theta }{\rho ^{2}}%
\right) \left[ dt+\frac{2aMr\sin ^{2}\theta }{\tilde{\Delta}-a^{2}\sin
^{2}\theta }d\phi \right] ^{2}-\frac{\rho ^{2}}{\tilde{\Delta}}\left[ dr^{2}+%
\tilde{\Delta}d\theta ^{2}\right] -\left[ \frac{\tilde{\Delta}\rho ^{2}\sin
^{2}\theta }{\tilde{\Delta}-a^{2}\sin ^{2}\theta }\right] d\phi ^{2},
\end{equation}

where $\rho ^{2}=r^{2}+a^{2}\cos ^{2}\theta ,$ and the constants $M$ and $a$
represents the emergent parameters in the local isometry for mass and
rotation, respectively. The roots of $\tilde{\Delta}$, namely $r_{+}$ and $%
r_{-}$ are the event (outer) and Cauchy (inner) horizons, respectively. From
this transformation, we conclude that the region of interaction is locally
isometric to the region in between the inner and outer horizon of the Kerr
black hole. The corresponding Weyl scalar in Boyer - Lindquist coordinates
becomes,%
\begin{equation}
\Psi _{2}=-\frac{M}{\left( r-ia\cos \theta \right) ^{3}}.
\end{equation}%
Figure 2, illustrates the region which is identical both in Kerr black hole
and in the interaction region of the CGW.

One may naturally look for a corresponding isometric colliding wave solution
belonging to the overspinning case $a>M$. This solution has not been
studied, however, the study which may be adopted to achieve such a solution
has been considered in \cite{xp}. In this paper, the prolate coordinate
system is used to obtain vacuum cylindrical gravitational waves. In
obtaining the solution, the Ernst formalism was employed. The solution to
the Ernst equation is obtained for $\xi ^{-1},$ instead of $\xi $ $\left(
=p\eta +iq\mu \right) $ and the condition $q^{2}-p^{2}=1,$ instead of $%
p^{2}+q^{2}=1.$ With this formalism it is shown in \cite{xp} that, the
overspinning Kerr solution can be derived via complex transformation from a
class of cylindrical wave spacetime. The scope of \cite{xp} was not to
derive a corresponding colliding gravitational wave solution but to connect
overspinning Kerr with a cylindrical wave spacetime. The fact that $%
q^{2}-p^{2}=1$ yields hyperbolic functions for $\eta $ and $\mu $ is
inappropriate for the colliding wave problem. Assuming that such functions
are expressed in the null coordinates $u$ and $v,$ the boundary conditions
for the incoming regions will not be satisfied across the null boundaries.
This naturally give rise to Dirac delta function sources at the boundaries
which is not acceptable for a vacuum problem. In short, correspondence of
overspinning Kerr solution to a CGW problem remains open.

\section{Quantum probes of timelike Kerr naked singularity}

In this section we seek to answer the following question: Is the only
classical singularity of Kerr, namely, $r=0$ and $\theta =\pi /2,$ also
quantum singular ?.

In Boyer - Lindquist coordinates $\left( t,r,\theta ,\phi \right) ,$ the
Kerr metric can be written as,%
\begin{equation}
ds^{2}=-\frac{\tilde{\Delta}}{\rho ^{2}}\left[ dt-a\sin ^{2}\theta d\phi %
\right] ^{2}+\frac{\rho ^{2}}{\tilde{\Delta}}dr^{2}+\rho ^{2}d\theta ^{2}+%
\frac{\sin ^{2}\theta }{\rho ^{2}}\left[ adt-\left( r^{2}-a^{2}\right) d\phi %
\right] ^{2},
\end{equation}%
note that the signature of the metric is changed to $+2$. If the rotational
parameter dominates the mass parameter (over spinning case, $a>M$), there
are no horizons and the timelike naked singularity at $r=0$ and $\theta =\pi
/2$ is developed for asymptotic observers. The metric (92) for the
particular case of $M=1$ is reduced to the following form, in the equatorial
plane $\theta =\pi /2$, 
\begin{equation}
ds^{2}=-\left( 1-\frac{2}{r}\right) dt^{2}-\frac{4a}{r}dtd\phi +\frac{r^{2}}{%
\bar{\Delta}}dr^{2}+r^{2}d\theta ^{2}+\left( r^{2}+a^{2}+\frac{2a^{2}}{r}%
\right) d\phi ^{2},
\end{equation}%
in which $\bar{\Delta}=r^{2}+a^{2}-2r,$ and the ranges of the coordinates
vary as%
\begin{equation}
0\leq r\leq \infty ,\text{ \ \ \ \ \ \ }0\leq \theta \leq \pi ,\text{ \ \ \
\ \ \ \ }0\leq \phi <2\pi .
\end{equation}%
In this case, the topology of the Kerr metric changes from ergospheres to
ergo torus in which the inner circle is the ring singularity and has a
timelike character (see \cite{grf} for figures). The timelike Kerr naked
singularity can be explained much better if we switched the coordinates to
the so-called Kerr - Schild \cite{ks} form in which the structure of $r=0$
surface becomes more transparent. The so-called Kerr - Schild forms $%
(t,x,y,z)$ are defined by%
\begin{equation}
x=\sqrt{r^{2}+a^{2}}\sin \theta \cos \left( \phi +\arctan \frac{a}{r}\right)
,
\end{equation}%
\begin{equation}
y=\sqrt{r^{2}+a^{2}}\sin \theta \sin \left( \phi +\arctan \frac{a}{r}\right)
,
\end{equation}%
\begin{equation}
z=r\cos \theta .
\end{equation}%
We have%
\begin{equation}
x^{2}+y^{2}+z^{2}=\left( r^{2}+a^{2}\right) \sin ^{2}\theta +r^{2}\cos
^{2}\theta .
\end{equation}%
Since the Weyl scalar (91) diverges at $r=0$ and $\theta =\pi /2,$ this
implies%
\begin{equation}
x^{2}+y^{2}=a^{2}\text{ \ \ \ \ \ \ \ \ \ at \ \ \ \ \ \ \ \ }z=0.
\end{equation}%
This is in fact a ring that forms the boundary of a disc. Hence, the only
singularity of the Kerr spacetime is located along this ring and the
interior of the ring $x^{2}+y^{2}<a^{2}$ remains regular.

In classical general relativity, the formation of a naked singularity can be
attributed as a threat to the cosmic censorship hypothesis of Penrose,
because, all singularities of gravitational collapse must be hidden within
black holes. Hence, the resolution of naked singularities constitute one of
the unresolved problem of black hole physics. It is believed that, the well
established quantum theory of gravity will be a powerful tool for the
resolution, however, the complete quantum theory of gravity is still under
"construction". In the literature, there are alternative methods for this
purpose. Loop quantum gravity \cite{ash} and string theory \cite{hr,nat} are
the two important study field in the resolution of singularities.

In this section, another alternative method will be used. The formation of
timelike naked singularities in the fast rotating case will be investigated
in view of quantum mechanics. Instead of point particle probes which leads
to the notion of \textit{geodesics incompleteness}, the wave probe will be
used which leads to the notion of \textit{quantum singularity}. In doing
this, the work of Wald \cite{wald} which was developed by Horowitz and
Marolf (HM) \cite{hm} for static spacetimes is extended to stationary
metrics. However, this extension is not a general extension. The main theme
of the HM criterion is to split the spatial and time part of the Klein-
Gordon equation and write it in the form of%
\begin{equation}
\frac{\partial ^{2}\psi }{\partial t^{2}}=-A\psi ,
\end{equation}%
where $A$ is the spatial wave operator. Note that, this operator is a
symmetric and positive operator on the Hilbert space $\mathcal{H}.$
According to the HM, the singular character of the spacetime with respect to
wave probe is characterized by investigating whether the spatial part of the
wave operator $A$ has a unique self - adjoint extensions (i.e. essentially
self - adjoint) in the entire space or not. If the extension is unique, it
is said that the space is quantum mechanically regular. In order to make
this point more clear, consider the Klein- Gordon equation for a free
particle that satisfies%
\begin{equation}
i\frac{d\psi }{dt}=\sqrt{A_{E}}\psi ,
\end{equation}%
whose solution is%
\begin{equation}
\psi \left( t\right) =e^{-it\sqrt{A_{E}}}\psi \left( 0\right) ,
\end{equation}%
in which $A_{E}$ denotes the extension of the operator $A$. If $A$ has not a
unique self - adjoint extensions, then the future time evolution of the wave
function (102) is ambiguous. And, HM criterion defines the spacetime as
quantum mechanically singular .(see \cite{hos}, for a detailed mathematical
background).

The timelike naked singularity for the Kerr metric will be probed with
scalar waves satisfying the Klein-Gordon equation%
\begin{equation}
\left( \frac{1}{\sqrt{g}}\partial _{\mu }\left[ \sqrt{g}g^{\mu \nu }\partial
_{\nu }\right] -\tilde{m}^{2}\right) \psi =0,
\end{equation}%
in which $\tilde{m}$ is the mass of the scalar particle. The considered
model of solution to the equation (103) is called a reduced wave equation
which admits solution in the form:%
\begin{equation}
\psi \left( t,r,\theta ,\phi \right) =e^{i\left( k\phi \right) }f(t,r,\theta
).
\end{equation}%
This form of choice for the solution of the Klein-Gordon equation is also
considered in \cite{bey}, for exploring the new symmetries of the solution
of the wave equation. For the metric (93), the Klein- Gordon equation with
the assumed solution can be written as%
\begin{equation}
\frac{\partial ^{2}f}{\partial t^{2}}+\frac{4aki}{r\Xi }\frac{\partial f}{%
\partial t}=\frac{\bar{\Delta}}{r^{2}\Xi }\left\{ \frac{\partial }{\partial r%
}\left( \bar{\Delta}\frac{\partial }{\partial r}\right) +\frac{\partial ^{2}%
}{\partial \theta ^{2}}-\left[ k^{2}\left( 1-\frac{a^{2}}{\bar{\Delta}}%
\right) +\tilde{m}^{2}r^{2}\right] \right\} f,
\end{equation}%
in which $f=f(t,r,\theta )$ and $k$ can take values of all integers which is
associated with the orbital quantum number and $\Xi =r^{2}+a^{2}+\frac{2a^{2}%
}{r}.$ The present form of equation (105), is not suitable to use because
the temporal and spatial parts are not yet seperable. The second term on the
left hand side of the equation (105) arose due to the presence of $g_{t\phi
} $ term in the metric (93). However, it is crucial to know that the formed
timelike naked singularity of the Kerr metric has some interesting
properties that is not shared by the any other naked singularities forming
in static spacetimes. The Kerr naked singularity becomes visible to
asymptotic observers, if one approaches to the singularity $r=0$ from $%
\theta =\pi /2$ only. In other words, the surface $r=0,$ is a disc with a
boundary of a ring singularity. The trajectories that approach to this
surface $r=0$ with $\theta \neq \pi /2,$ do not fall into the singularity,
and hence, all points are regular.

In the assumed solution in equation (104), the constant parameter $k$ which
runs for all integer values is related to the orbital quantum number
corresponding to the projection of the angular momentum onto the axis of
symmetry. In order to probe the Kerr naked singularity with waves, the wave
should propagate along the equilateral plane $\theta =\pi /2.$ In addition
to this, the ring singularity is located at $r=0$ surface. These
restrictions on the wave propagation imposes the condition that the only
wave mode available for this probe is the \textit{s-wave} mode that
corresponds to $k=0$. This coincidence enables the equation (105) separable
in time and spatial part as

\begin{equation}
\frac{\partial ^{2}f}{\partial t^{2}}=\frac{\bar{\Delta}}{r^{2}\Xi }\left\{ 
\frac{\partial }{\partial r}\left( \bar{\Delta}\frac{\partial }{\partial r}%
\right) +\frac{\partial ^{2}}{\partial \theta ^{2}}-\tilde{m}%
^{2}r^{2}\right\} f,
\end{equation}%
and the spatial wave operator $A$ which will be investigated for a unique
self - adjoint extension has a form of%
\begin{equation}
A=-\frac{\bar{\Delta}}{r^{2}\Xi }\left\{ \frac{\partial }{\partial r}\left( 
\bar{\Delta}\frac{\partial }{\partial r}\right) +\frac{\partial ^{2}}{%
\partial \theta ^{2}}-\tilde{m}^{2}r^{2}\right\} .
\end{equation}

The problem now is to count the number of extensions of the operator $A$.
This is done by using the concept of deficiency indices discovered by Weyl 
\cite{wyl} and generalized by von Neumann \cite{vnm} (see \cite{hos} for a
detailed mathematical background). The determination of the deficiency
indices $\left( n_{+},n_{-}\right) $ of the operator $A$, is reduced to
count the number of solutions to equation%
\begin{equation}
A\psi \pm i\psi =0,
\end{equation}%
that belong to the Hilbert space $\mathcal{H}.$ If there are no square
integrable ($L^{2}\left( 0,\infty \right) $) solutions (i.e., $n_{+}=n_{-}=0$%
) in the entire space, the operator $A$ possesses a unique self-adjoint
extension and it is called essentially self-adjoint. Consequently, the
method to find a sufficient condition for the operator $A$ to be essentially
self-adjoint is to investigate the solutions satisfying equation (108)\ that
do not belong to the Hilbert space $\mathcal{H}$.

The solution to Eq.(108) is obtained by assuming the solution in separable
form \ $\psi =R\left( r\right) Y\left( \theta \right) ,$ which yields the
radial equation as

\begin{equation}
R^{\prime \prime }+\frac{2\left( r-1\right) }{\bar{\Delta}}R^{\prime }-\frac{%
1}{\bar{\Delta}}\left[ r^{2}\left( \tilde{m}^{2}\pm \frac{i\Xi }{\bar{\Delta}%
}\right) +c\right] R=0,
\end{equation}%
in which prime denotes the derivative with respect to $r$ and $c$ is a real
separation constant.

The square integrability of the solutions of Eq.(109) for each sign $\pm $
is checked by calculating the squared norm of the solution of Eq.(109) for
the massless wave $\tilde{m}=0$, in which the function space on each $t=$
constant hypersurface $\Sigma _{t}$ is defined as $\mathcal{H}=\left\{ R\mid
\parallel R\parallel <\infty \right\} .$ The squared norm can be defined as 
\cite{hm},%
\begin{equation}
\Vert R\Vert ^{2}=\int_{\Sigma _{t}}\sqrt{-g}g^{tt}RR^{\ast }d^{3}\Sigma
_{t}.
\end{equation}%
The spatial operator $A$ is essentially self-adjoint if neither of the
solutions of Eq.(109) is square integrable over all space $L^{2}\left(
0,\infty \right) .$ The behavior of the Eq.(109) near $r\rightarrow 0$ and $%
r\rightarrow \infty $ will be considered separately in the following
subsections.

\subsection{The case of $r\rightarrow 0:$}

In the case when $r\rightarrow 0,$ the Eq.(109) simplifies to,

\begin{equation}
\psi ^{\prime \prime }-\frac{2}{a^{2}}\psi ^{\prime }-\frac{c}{a^{2}}\psi =0.
\end{equation}%
If the separation constant $c>-\frac{1}{a^{2}}$, then the solution is%
\begin{equation}
R(r)=e^{r/a^{2}}\left( C_{1}e^{\alpha _{1}r/a^{2}}+C_{2}e^{-\alpha
_{1}r/a^{2}}\right) ,
\end{equation}%
in which $\alpha _{1}=\sqrt{1+ca^{2}}.$ If the separation constant $c<-\frac{%
1}{a^{2}},$ then $\alpha _{1}\rightarrow i\beta _{1},$ where $\beta _{1}$ is
an arbitrary real constant and the solution is%
\begin{equation}
R(r)=e^{r/a^{2}}\left( C_{3}e^{i\beta _{1}r/a^{2}}+C_{4}e^{-i\beta
_{1}r/a^{2}}\right) ,
\end{equation}%
in which $C_{i}$ $\left( i=1,2,3,4\right) $ are the integration constants.

The square integrability of the solution (112) and (113) are checked by
calculating the squared norm defined in equation (110) in the limiting case
of the metric (93) when $r\rightarrow 0,$ which is given by%
\begin{equation}
\Vert R\Vert ^{2}\sim \int_{0}^{const.}r^{5/2}\left\vert R\right\vert ^{2}dr.
\end{equation}%
The analysis has revealed that, if the separation constant $c<-\frac{1}{a^{2}%
}$, the solution (113) is square integrable, since $\Vert R\Vert ^{2}<\infty
,$ thus, the solution belongs to the Hilbert space. But, there is a specific
case for $c>-\frac{1}{a^{2}}$ in the solution (112) such that, if the
separation constant is chosen very large$,$ then, this specific solution
fails to satisfy square integrability condition, i.e. $\Vert R\Vert
^{2}\rightarrow \infty .$

\subsection{The case of $r\rightarrow \infty :$}

When $r\rightarrow \infty ,$ the Eq.(109) reduces to%
\begin{equation}
R^{\prime \prime }+\frac{2}{r}R^{\prime }\pm iR=0,
\end{equation}%
whose solution is given by%
\begin{equation}
R(r)=\frac{1}{r}\left( C_{3}\sin \left( \alpha _{2}r\right) +C_{4}\cos
\left( \alpha _{2}r\right) \right) ,
\end{equation}%
in which $\alpha _{2}=\frac{1}{2}\sqrt{\pm 1+i},$ and $C_{3}$ and $C_{4}$
are the integration constants. The square integrability is checked with the
following norm written for the case $r\rightarrow \infty ,$%
\begin{equation}
\Vert R\Vert ^{2}\sim \int_{const.}^{\infty }r^{2}\left\vert R\right\vert
^{2}dr.
\end{equation}%
The result is that the solution fails to satisfy square integrability
condition ($\Vert R\Vert ^{2}\rightarrow \infty $), and hence, does not
belong to the Hilbert space.

The method of defining whether the operator $A$ has a unique self-adjoint
extension (or essentially self-adjoint) is to investigate the solution of
Eq.(109) in the entire space $\left( 0,\infty \right) $ and count the number
of solutions that do not belong to the Hilbert space. In other words, if
there is one solution that fails to be square integrable for the entire
space then the operator $A$ is said to be essentially self-adjoint. Our
analysis has shown that the behaviour of Eq. (109), when $r\rightarrow
\infty ,$ admits solution that is not square integrable. Hence, the operator 
$A$ is essentially self-adjoint and the future time evolution of the quantum
particles/waves can be predicted uniquely. Consequently, the classical Kerr
naked ring singularity is healed and becomes quantum mechanically regular
when probed with particles/waves described by the Klein-Gordon equation.

We would like to emphasize that the criterion proposed by HM \cite{hm}, by
adopting the earlier work of Wald \cite{wald}, is used for probing the
timelike singularities in static spacetimes. A quantum mechanical particle
with mass $\tilde{m}$ can be described by the Klein - Gordon equation in the
form of Eq. (100). In this equation, there is a complete separation in the
time and spatial parts. Hence, the operator $A$ is completely defined in
terms of spatial coordinates and spatial derivatives only. This leads to
Eq.(102), that may be interpreted as, translating the well-posedness of the
initial value problem into the essential self-adjointness of the operator $A$
\cite{hos}. In this paper, the adressed problem is the Kerr timelike naked
ring singularity that developed in stationary spacetime. The Klein-Gordon
equation for this spacetime is given in Eq. (105). Since, complete
separation in the spatial and time derivatives is not possible, one may
define the right hand side of Eq. (105) as the spatial part of the reduced
normalized wave operator \cite{bey}, without imposing $k=0$ (\textit{s-wave}%
). However, even if we consider this case, the result would not change,
because, the constant number $k$ do not contribute near $r\rightarrow 0$ and 
$r\rightarrow \infty .$ As a result, we would obtain exactly the same
behaviour of the operator $A$ given in Eq.(106).

\section{Conclusion and Discussion}

The local isometry between a Kerr black hole and a CGW spacetime is known as
the CX duality. A CH forming CGW spacetime transforms locally by a
coordinate transformation into a black hole metric. Is this a coincidence ?.
Can such a duality be valid for all black holes / colliding shock waves ?. A
stable black hole from classical physics point of view is a highly
localized, simplest object in our universe. It is a long time, more than a
half century that stability of black holes is investigated. From quantum
physics point of view it was shown by Hawking in 1970's that black holes
undergo a thermal radiation. Such a radiation causes the black hole to
evaporate completely or leave behind a stable remnant. Let us remember also
that in a quantum theory, due to the uncertainty principle even exact
location of the horizon is questionable: a fuzzy picture becomes
indispensable at the Planck scale. Different classes of CGW such as KP and
NH form naked singularities instead of the analytically extendible \textit{%
"beautiful"} CH, so that an observer falls into the singularity without
crossing an event horizon. Further, the beauty of CH in CGW is destroyed by
the ugly fact: it is not stable against various perturbations so that it
settles down to a naked singularity. A naked singularity is known to violate
the cosmic censorship conjecture. For a diagonal metric such as
Schwarzschild, once the observer crosses the event horizon she can not
distinguish between a black hole and colliding wave spacetimes. ( Note that
for an off - diagonal black hole metric such as Kerr, there is an extra
region, the ergosphere, depicted in Fig. (2)). This is a lesson that we
learn from the CX duality. It is our belief that the key to understand
quantum gravity through the wave - particle duality lies in understanding
the event horizon. Direct experimental discovery of an event horizon either
in space or in analog gravity models in a laboratory will highlight a 
\textit{"miracle without miracle"} in John A. Wheeler's terminology. We may
anticipate, without a priori proof, that solutions to information paradox,
entanglement, complementarity, firewall and other mind boggling concepts can
be tackled all with a thorough understanding of the physics of horizons \cite%
{ms}. Let the Kerr geometry inspire / guide us more and more toward this
goal.

Finally, the formation of the timelike Kerr naked singularity in the
overspinning case is analyzed in view of quantum mechanics with the
criterion proposed by HM. This singularity is probed with waves obeying the
Klein-Gordon equation. Analysis has revealed that, in order for probing the
ring singularity, the spatial derivative operator of the reduced
Klein-Gordon equation is defined for a specific wave mode, namely the 
\textit{s-wave}. It is shown that the spatial derivative operator $A$
defined in Eq. (106) is essentially self-adjoint. As a result, the classical
Kerr ring singularity is quantum regular with respect to the wave probe
described by the reduced Klein-Gordon equation.

The wave mode used in this study is the only wave mode that makes it
possible to use the criterion proposed by HM. The general extension of the
criterion of HM for stationary spacetimes has not been achieved yet and this
problem is an open problem. Although, the preliminary work for stationary
spacetimes is considered in \cite{seg}, however, the formulation has not
been fully completed.

\textbf{Acknowledgment:}

We wish to thank S. H. Mazharimousavi and the anonymous referee for their
helpful comments. In particular, suggestions by the referee helped us to
improve the previous version of the paper.

\end{document}